\documentstyle[epsf]{aa}

\newcommand{\lsim}{\raisebox{0.3mm}{\em $\, <$} \hspace{-3.3mm} 
\raisebox{-1.8mm}{\em $\sim \,$}}

\def\etal{{\rm et al.\ }}
\def\vs{{\rm v.s.\ }}

\begin{document}

\thesaurus{03         
           (11.03.1;  
            11.05.1;  
            11.05.2;  
            11.06.1;  
            11.16.1;  
            11.19.5)} 
\title{Evolution of the colour-magnitude relation of early-type galaxies in
       distant clusters}


\author{Tadayuki Kodama\inst{1,2}
        \thanks{\emph{e-mail:} kodama@ast.cam.ac.uk}
        \and Nobuo Arimoto\inst{2,4}
        \and Amy J.\,Barger\inst{1,3}
        \and Alfonso Arag\'on-Salamanca\inst{1}}

\offprints{T. Kodama}

\institute{Institute of Astronomy, University of Cambridge,
           Madingley Road, Cambridge CB3 0HA, UK
           \and Institute of Astronomy, University of Tokyo, Mitaka,
           Tokyo 181, Japan
           \and Institute for Astronomy, University of Hawaii, 2680 Woodlawn
           Drive, Honolulu, Hawaii 96822, USA
           \and Observatoire de Paris, Section de Meudon, DAEC, 
           Meudon Principle Cedex, 92195, France}

\date{Received --- --, --- / Accepted --- --, ---}

\maketitle

\begin{abstract}

We present a thorough quantitative analysis of the evolution of the
colour-magnitude relation for early-type galaxies in 17 distant
clusters with redshifts $0.31 < z < 1.27$ using the Kodama \& Arimoto
(1997) evolutionary model for elliptical galaxies.  The model is
calibrated to reproduce the colour-magnitude relation for Coma
ellipticals at $z\sim 0$ and gives the evolution of the slope and
zero-point as a function of redshift. We find no significant
differences between the colour-magnitude relations of the clusters in
our sample.  The slopes can be reproduced by a single model sequence in
which all elliptical galaxies are assumed to be equally old (the
maximum age difference allowed for the brightest 3 magnitudes is only
1~Gyr) and to have mean stellar metallicities which vary as a function
of galaxy luminosity. The zero-points of the colour-magnitude relations
constrain the epoch of major star formation in early-type galaxies to
$z_f>2$--$4$.
This study provides two important constraints for any model of the
formation of rich clusters: the uniformity of the ages of the stellar
populations in the early-type galaxies and the universality of the
metallicity sequence of these galaxies as a function of galaxy mass.


\end{abstract}


\section{Introduction}

Visvanathan \& Sandage (1977) first noted from their data on 9 nearby
clusters that the more luminous early-type galaxies tended to have
redder colours. Bower, Lucey, \& Ellis (1992; hereafter BLE92)
later studied this {\it
colour-magnitude} ({\it C-M}) relation in detail using high precision
photometry of early-type galaxies in the Virgo and Coma clusters and
found very little scatter about the mean {\it C-M\/} relation.  Thus,
there appeared to be a marked homogeneity in the present-day early-type
cluster galaxy population.  Recently, Kodama \& Arimoto (1997;
hereafter KA97) compared the predictions of their evolutionary models
of elliptical galaxies with the {\it C-M\/} diagrams for early-type
galaxies in two distant clusters, Abell 2390 at $z=0.228$ and Abell 851
at $z=0.407$, and showed that the origin of the {\it C-M\/} relation
was primarily due to the mean stellar metallicity of the early-types
being a function of total magnitude.  In this paper we present a more
thorough study of the evolution of the {\it C-M\/} relation by using 17
clusters at cosmological distances to detect the colour evolution of
cluster early-type galaxies directly and then comparing the results
with the models. This will enable us to put even stronger constraints
on the formation of early-type galaxies in clusters.

The photometric evolution of early-type galaxies in $z<1$ clusters was
previously examined by several authors (e.g., Ellis et al.\ 1985; Couch,
Shanks \& Pence\ 1985; Arag\'on-Salamanca et al.\ 1993).
Arag\'on-Salamanca et al.\ (1993) traced the evolution of the
`red envelope' from the
optical-near-infrared colour distribution of galaxies in 10 clusters
with redshifts $0.5 < z < 0.9$ and concluded that the detected
evolution was consistent with the passive ageing of stellar populations
formed before $z\simeq 2$.  A potential problem with these studies is
that the early-type galaxies had to be selected by their spectral
energy distributions and colours, thereby increasing the possibility of
contamination from other galaxy types. The advent of the Hubble Space
Telescope {\it (HST)}, with its spectacular high-resolution images, has
made the morphological classification of distant galaxies possible.
Furthermore, thanks to the {\it HST}, large ground-based telescopes,
and the high quality of astronomical detectors, much fainter limiting
magnitudes can now be reached.  An additional advantage of {\it HST}
photometry comes from the reduction of the sky background at longer
wavelengths compared to that achieved with ground-based observations
(e.g. the background in space is 8 times fainter in the $I$-band),
thereby reducing the photometric errors at faint limits significantly
(Ellis et al.\ 1997; hereafter E97).  
These improvements now make it possible to
examine the evolution of early-type galaxies over a much wider
luminosity and redshift range.

The {\it C-M\/} relation is a powerful tool for quantifying the colour
evolution of early-type galaxies as a function of redshift since it can
be observed up to very high redshifts.  Dickinson (1996) showed that
the {\it C-M\/} relation is already recognizable at $z\simeq 1.2$.  The
zero-point of the {\it C-M\/} relation at high redshift provides direct
information on the properties of early-type galaxies in general.  On
the other hand, the slope of the {\it C-M\/} relation has information
on differential properties of early-type galaxies as a function of
luminosity or galaxy size.

Stanford, Eisenhardt, \& Dickinson (1995; 1998, hereafter SED98)
presented {\it C-M\/} relations for morphologically selected early-type
galaxies in 19 clusters out to $z=0.9$. When they compared their
optical-near-infrared {\it C-M\/} relations to that of Coma, they found
no significant change in either the slope or the scatter as a function
of redshift. They did, however, observe a progressive blueing with
redshift of the average colour in a manner consistent with the passive
evolution of an old stellar population formed in a burst at an early
cosmic epoch.

E97 also analyzed the {\it C-M\/}
relations of three distant clusters at $z\simeq 0.54$ (Cl~0016$+$16, Cl~
0054$-$27, Cl~0412$-$65), in this case using {\it HST} photometry.
They found that the dispersion about the relations was quite small even
in such distant clusters, thereby requiring a high formation redshift
for the stars in early-type galaxies, such as $z\simeq3$. E97 also did
not detect a significant change in the slopes of the relations in the
rest-frame $U-V$ when compared with Coma.  The zero-points show modest
colour evolution in agreement with earlier studies (Ellis et al. 1985;
Couch et al. 1985; Arag\'on-Salamanca et al. 1993).  All of these
results support a formation picture of early-type galaxies in which the
bulk of the star formation is completed at high redshift with little
star formation occurring in the recent past.  Thus, the {\it C-M\/}
relation originates at high redshift as a metallicity effect.

Recently, however, Kauffmann \& Charlot (1997) have argued that the
{\it C-M\/} relation can alternatively be interpreted in the context of
the hierarchical models of galaxy formation once chemical enrichment is
taken into account.  In their analysis, the slope of the {\it C-M\/}
relation is maintained because large ellipticals form primarily from
large metal-rich progenitors. They find that it remains nearly constant
up to $z\simeq 1$. Furthermore, they are able to explain the tightness
of the {\it C-M\/} relation, despite frequent galaxy merging, by
suggesting that we are biasing ourselves to the selection of only old
galaxies by studying rich clusters at high redshift. According to their
models, these objects are not the progenitors of present-day clusters.
It is not yet clear whether this type of model can give an universal
{\it C-M\/} relation for different clusters, since merging histories
are likely to have varied from cluster to cluster.

Our motivations to conduct further analyses on the evolution of the
{\it C-M\/} relation for cluster early-type galaxies are twofold:
(1)~to investigate the universality of the {\it C-M\/} relation in
clusters and (2)~to put stronger constraints on the formation of
early-type galaxies by using the more realistic elliptical galaxy model
of KA97, which takes into account chemical evolution, rather than an
ad-hoc single burst model with fixed metallicity ($Z_\odot$) adopted by
most of the previous analyses.  KA97 have constructed an evolutionary
model of elliptical galaxies which uses a population synthesis
technique based on a monolithic collapse picture of galaxy formation.
The marked advantage of the KA97 model is that it allows us to analyze
both the slope of the {\it C-M\/} relation and the zero-point, since it
includes the effects of metallicity.  Moreover, since the model is
calibrated with the empirical colours of ellipticals in Coma, it can be
directly and reliably compared to the observational data in distant
clusters.

In this paper, we investigate the evolution of the {\it C-M\/} relation
of 17 distant clusters using the KA97 evolutionary models.  We have
accumulated the photometric data from the literature, most of which
were obtained with the {\it HST} and large, ground-based telescopes
($\sim 2$--$4\,$m). Although some of the {\it HST} data suffer from
zero-point uncertainties due to the paucity of observed standard stars
(see below), their random photometric errors are very small down to
$\sim 3\,$mag from the brightest end of the {\it C-M\/} relation
(typically $0.01-0.07\,$mag). Thus, we are able to conduct a reliable
analysis of, for example, the slope of the {\it C-M\/} relation based
on the relative photometry within each cluster.  To constrain the
formation epoch of early-type galaxies in clusters, we also conduct a
zero-point analysis using only those clusters which have been
calibrated with ground-based photometry or those which are at
sufficiently high redshift that the evolutionary changes are large
enough to put some constraint on the formation epoch, despite
relatively large zero-point uncertainties.

The cosmological parameters we have chosen to use throughout this paper
are $H_0=50\,$km$\,$s$^{-1}\,$Mpc$^{-1}$ ($h\equiv H_0/100 = 0.5$) and
$q_0=0.5$, without a lambda term, unless otherwise stated.
The structure of this paper is as follows. In \S2 we compile the
observational data and define the observed {\it C-M\/} relation for
each cluster. In \S3 we give a brief description of the evolutionary
models of KA97. We compare the models and the observations in \S4, and
in \S5 we give a discussion of the results and our conclusions.

\section{Data}

\begin{table*}
\caption{Cluster samples and the regression lines for the {\it C-M\/} relations.}
\label{table-1}
\begin{center}
  \begin{tabular}{llrrclllccc}
  \hline
   cluster      & $z$   & N$_{1}$ & N$_{2}$ & d.a. & colour & mag &
                $M_{0}$ & $A$ & $B$ & ref\\
  \hline
   AC~103       & 0.31  & 25 &  4 & 28 & $R_{702}-K$       & $K_{\rm rest}$ &
                $-$25 & $3.23 \pm 0.04$ & $-0.047 \pm 0.065$ & 1\\
                &       & 25 &  4 & 28 & $I-K$             & $K_{\rm rest}$ &
                $-$25 & $2.70 \pm 0.04$ & $-0.054 \pm 0.052$ & 1\\
   AC~114       & 0.31  & 30 &  4 & 28 & $R_{702}-K$       & $K_{\rm rest}$ &
                $-$25 & $3.13 \pm 0.03$ & $-0.027 \pm 0.032$ & 1\\
                &       & 30 &  4 & 28 & $I-K$             & $K_{\rm rest}$ &
                $-$25 & $3.33 \pm 0.02$ & $-0.083 \pm 0.024$ & 1\\
   AC~118       & 0.31  & 35 &  3 & 28 & $R_{702}-K$       & $K_{\rm rest}$ &
                $-$25 & $3.04 \pm 0.03$ & $-0.115 \pm 0.027$ & 1\\
                &       & 35 &  4 & 28 & $I-K$             & $K_{\rm rest}$ &
                $-$25 & $2.77 \pm 0.02$ & $-0.134 \pm 0.028$ & 1\\
   Cl~$0024+16$ & 0.39  & 28 &  0 & 32 & $I_{814}-K$       & $K_{\rm rest}$ &
                $-$25 & $3.08 \pm 0.02$ & $-0.091 \pm 0.028$ & 1\\
 3C~295         & 0.461 & 25 & --- & 24 & $V-K$            & $K^T$          &
                $-$25.5& $4.98 \pm 0.07$ & $-0.09 \pm 0.04$  & 3\\
   Cl~$0412-65$ & 0.510 & 29 &  9 & 10 & $V_{555}-I_{814}$ & $I_{814}^{T}$  &
                $-$22 & $2.32 \pm 0.03$ & $-0.091 \pm 0.029$ & 2\\
                &       & 29 & --- & 24 & $V-K$            & $K^T$          &
                $-$25.5& $5.30 \pm 0.06$ & $-0.09 \pm 0.04$  & 3\\
 GHO~$1601+4253$& 0.539 & 42 & --- & 26 & $V-K$            & $K^T$          &
                $-$25.5& $5.33 \pm 0.06$ & $-0.09 \pm 0.04$  & 3\\
MS~$0451.6-0306$& 0.539 & 51 & --- & 26 & $V-K$            & $K^T$          &
                $-$25.5& $5.24 \pm 0.06$ & $-0.12 \pm 0.03$  & 3\\
   Cl~$0016+16$ & 0.546 & 25 &  4 & 36 & $I_{814}-K$       & $K_{\rm rest}$ &
                $-$25 & $2.87 \pm 0.02$ & $-0.046 \pm 0.031$ & 1\\
                &       &106 & 15 & 10 & $V_{555}-I_{814}$ & $I_{814}^{T}$  &
                $-$22 & $2.38 \pm 0.01$ & $-0.067 \pm 0.010$ & 2\\
                &       & 65 & --- & 26 & $V-K$            & $K^T$          &
                $-$25.5& $5.20 \pm 0.07$ & $-0.08 \pm 0.02$  & 3\\
   Cl~$0054-27$ & 0.563 & 36 &  6 & 10 & $V_{555}-I_{814}$ & $I_{814}^{T}$  &
                $-$22 & $2.28 \pm 0.02$ & $-0.090 \pm 0.015$ & 2\\
                &       & 38 & --- & 26 & $V-K$            & $K^T$          &
                $-$25.5& $5.34 \pm 0.06$ & $-0.14 \pm 0.03$  & 3\\
 3C~220.1       & 0.620 & 22 & --- & 27 & $V-K$            & $K^T$          &
                $-$25.5& $5.62 \pm 0.07$ & $-0.13 \pm 0.04$  & 3\\
 3C~34          & 0.689 & 19 & --- & 24 & $V-K$            & $K^T$          &
                $-$25.5& $5.78 \pm 0.06$ & $-0.11 \pm 0.04$  & 3\\
 GHO~$1322+3027$& 0.751 & 23 & --- & 24 & $R-K$            & $K^T$          &
                $-$26  & $4.56 \pm 0.06$ & $-0.07 \pm 0.03$  & 3\\
 MS~$1054.5-032$& 0.828 & 71 & --- & 22 & $R-K$            & $K^T$          &
                $-$26  & $4.85 \pm 0.06$ & $-0.11 \pm 0.02$  & 3\\
 GHO~$1603+4313$& 0.895 & 23 & --- & 21 & $R-K$            & $K^T$          &
                $-$26  & $5.01 \pm 0.06$ & $-0.13 \pm 0.03$  & 3\\
   3C~324       & 1.206 & 13 &  1 & ---& $R-K$             & $K$            &
                $-$26 & $5.84 \pm 0.05$ & $-0.099 \pm 0.065$ & 4\\
CIG~J$0848+4453$& 1.273 &  6 &  0 & 17 & $R-K$             & $K$            &
                      & $^{\dagger}$$5.92 \pm 0.25$ &                    & 5\\
  \hline
  \end{tabular}\\
  \small
N$_1$ --- total number of ellipticals, \ N$_2$ --- number of excluded objects, \ d.a. --- diameter aperture in kpc,\\
$M_0$ --- absolute magnitude of C-M zero-point, \ $A$ and $B$ --- coefficients of C-M relation ($^{\dagger}$ shows average colour).
,\\
refs. --- 1: Barger et al. (1997), 2: Ellis et al. (1997),
          3: Stanford et al. (1998), 4: Dickinson (1996),
          5: Stanford  et al. (1997)\\
   \end{center}
\end{table*}

\subsection{Sources}

Our dataset for this analysis
consists of 16 clusters with redshifts between $z=0.31$ and $z=1.206$ 
from various sources, all of which have been imaged
with the Wide Field and Planetary Camera 2 (WFPC2) on the {\it HST}. Here
we denote the {\it HST} magnitudes in the WFPC2 filters 
F555W, F702W, and F814W by $V_{555}$, $R_{702}$, and $I_{814}$, respectively. 
We have included one additional cluster,
CIG~J0848$+$4453 at $z=1.273$, which was not observed with the {\it HST},
but being at the highest redshift, it can place useful constraints. 

In this paper we construct optical-$K$ {\it C-M\/} relations for six
clusters, namely AC~103, AC~114, AC~118, Cl~0024$+$16, Cl~0016$+$16,
and 3C~324. For the remaining clusters we adopt {\it C-M\/} relations
given in the literature:  (1) ground-based optical$-K$ colour versus
total $K$ magnitude for 11 clusters between $z=0.461$ and 0.895 from
SED98; (2) {\it HST} optical-optical ($V_{555}-I_{814}$) colour versus
total $I_{814}$ magnitude for 3 clusters at $z\sim 0.55$ from E97.

A complete list of the clusters used in this study are given in
Table~1, along with a summary of their relevant properties.  Columns
1--4 list the cluster name, redshift, total number of isolated,
morphologically classified spheroidals within the magnitude limit, and
the total number of spheroidals excluded from the analysis because of
their anomalous colours (see discussion below).  Column~5 gives the
adopted apertures in which the colour and magnitude measurements were
made, in units of kpc.  Note that the $I_{814}$ magnitudes from E97 and
the $K$ magnitudes from SED98 are total magnitudes, as indicated by the
subscript $T$.  The colours and magnitudes used to define the {\it C-M\/}
relation for each cluster are given in Columns~6 and 7, respectively.
The subscript `rest' means the magnitude is in the rest frame; all
others are given in the frame of the observer.  Column~8 lists the
absolute magnitude, $M_{0}$, which defines the zero-point of the {\it
C-M} relation for each cluster, and Columns~9 and 10 give the
corresponding {\it C-M} relation regression coefficients (see \S2.2).
Finally, Column~11 lists the references for the data.

\subsection{Colour-magnitude relations}

For five clusters, AC~103, AC~114, AC~118, Cl~0024$+$16, and
Cl~0016$+$16, we have used the morphological classifications given in
Couch et al.\ (1997) and Smail et al.\ (1997) to select a sample of
{\it isolated} (i.e. uncontaminated by nearby objects which would
distort the colours) early-type (E, E/S0, or S0) galaxies from the
cluster cores.

The reliability of the morphological selection of spheroidal galaxies
in clusters at $z\sim 0.55$ was examined in detail by E97.  Visual
classifications were found to be robust to $I_{814}\le 21.0$, but over
the interval $I_{814}=21.0 - 22.0$ the distinction between the
early-type classes became increasingly uncertain. Principally for this
reason we have decided not to attempt to separately analyze the
elliptical and S0 galaxy populations. This will also increase our
sample size and hence improve the statistics. Since E97 found no
evidence for a distinction between the {\it C-M\/} relations for the E
and S0 populations, our decision to combine these morphological types
should not affect the results of our analysis.  Moreover, Dressler et
al. (1997) point out that the number fraction of S0 galaxies in
clusters are rapidly decreasing as a function of redshift, and above
redshift 0.6 or so, S0 galaxies could almost disappear.

To construct our {\it C-M\/} diagrams,
we use ground-based $I$ and $K$-band data for AC~103, AC~114 and AC~118 
from Barger et al.\ (1996) and ground-based $K$-band data for Cl~0024$+$16 and
Cl~0016$+$16 from Barger et al.\ (1997). We also use existing {\it HST} 
optical data in either the $R_{702}$ or $I_{814}$ bands. 
We selected magnitude limits such that the samples
would contain approximately the brightest 3$\,$mag in the $K$ band. 
The actual absolute magnitude limits depend on the depth of
the $K$-band images and the redshift of the clusters, but they all
lie between $M_{K}=-23$ and $-24$.
We present $K$-band magnitudes as absolute magnitudes 
in the rest frame for the clusters in our sub-sample.
The $k$-corrections are determined from a typical spectral energy
distribution (SED) of giant elliptical galaxies (Arag\'on-Salamamca et
al. 1993).  The SED changes along the {\it C-M\/} relation and hence the
correct amount of $k$-correction is varied towards fainter galaxies,
however this effect is negligible. Acording to the model SEDs, the
difference in the $k$-correction across the {\it C-M\/} relation is about
only 0.1 magnitude at most, and the effect on both the {\it C-M\/} slopes
and the zero-points are negligible in our analysis.
The seeing corrections are made from an estimate of how much light falls
outside the adopted aperture due to scattering, and the Galactic extinction
corrections are determined from the $E(B-V)$ values toward the clusters 
as estimated by the HI intensity (Burstein \& Heiles 1984).
We adopt $E(B-V)=0.04$ for AC~114, 0.05 for AC~103, 0.055 for AC~118,
and 0.03 for Cl~0024$+$16 and Cl~0016$+$16.

For the high redshift cluster 3C~324 at $z=1.206$, we adopted $R-K$
colours and $K$ magnitudes of 13 galaxies in the `red finger' with
$K<19\,$mag (Dickinson 1996).  These galaxies have been confirmed to be
early-types from {\it HST\/} WFPC2 imaging.

\begin{figure*}
\begin{center}
  \leavevmode
  \epsfxsize 1.0\hsize
  \epsffile{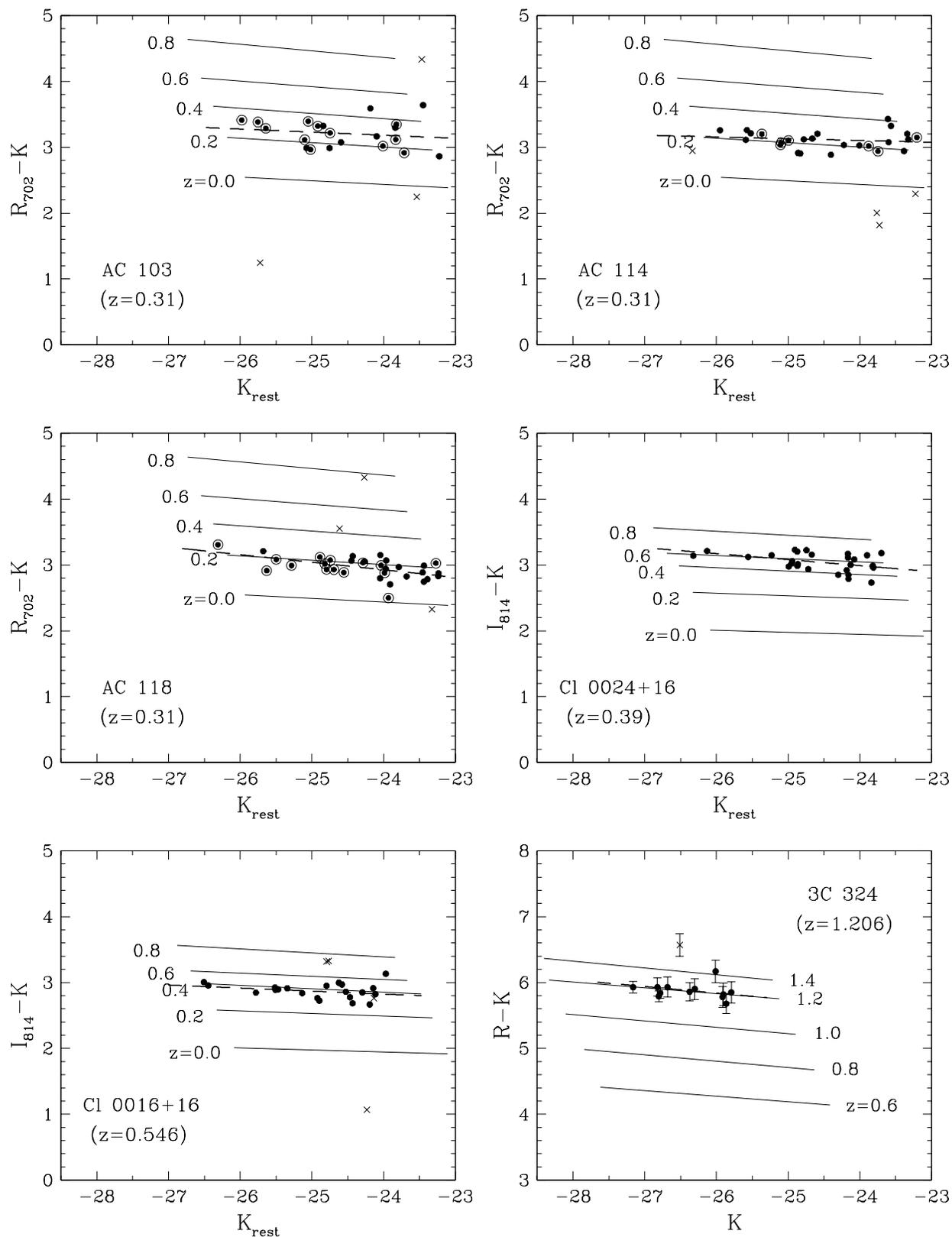}
\end{center}
\vspace{-0.5cm}
\caption
{
 {\it C-M} diagrams for the spheroidal galaxies in 6 of the clusters in our
 sample.
 Filled circles represent the spheroidal population in each cluster.
 An open circle surrounding a filled dot represents a spectroscopically
 confirmed cluster member. A cross indicates an apparent non-member, which
 we have excluded
 from our analysis. The dashed lines show the {\it C-M\/} relations as defined
 in the text.
 The solid lines show the model with $T_G=12$ Gyr ($z_f\simeq4.5$).
 The redshifts of the models increase from bottom to top, as indicated. 
}
\label{fig:diagram}
\end{figure*}

Figure~\ref{fig:diagram} shows the {\it C-M\/} diagrams for the
early-type galaxies in our sub-sample of six clusters above.
The open circles with filled dots represent the spectroscopically
confirmed members. The crosses indicate the objects whose colours place
them clearly outside the {\it C-M\/} relation.
Although there are very few such objects in our sample (see N$_2$ in
Table~1), we have excluded them from our analysis since their anomalous 
colours could distort the global statistics.
Most of these excluded objects are likely to be non-members,
but it is possible that a fraction of the bluer objects 
are members that have recently undergone strong star formation.
However, these excluded objects only account for a very small fraction
(i.e. 5~\%) of the total stellar mass in the early-type galaxies
in our six sub-sample clusters (estimated from the $K$-band luminosity),
hence they will not affect our analysis of the bulk of the stellar population.
The remainder of the galaxies are represented with
filled circles.  Photometric random errors are shown only for 3C~324.
All the other clusters have negligibly small errors, smaller than the
size of the symbols.

We calculate the {\it C-M\/} regression lines using the BCES
(bivariate, correlate errors and scatter) method of Akritas \& Bershady
(1996), who have kindly provided us with their software. Their
program gives both the regression lines and the fitting errors
analytically, considering the errors in both the magnitudes and the
colours, as well as the intrinsic scatter of the data.  The covariant
errors are assumed to be equal to the errors in the magnitudes since
the magnitude errors dominate the errors in the colours.  Because the
program considers the intrinsic scatter of the data, the error
estimations of the fitting parameters are reliable; otherwise the
fitting errors would be greatly underestimated.  The regression line
fits to the data are illustrated with the dashed lines in each of the
{\it C-M\/} diagrams of Fig.~\ref{fig:diagram}.

The parameter fits are summarized in Table~1, where
`$A$' is the zero-point colour of the regression line at the
absolute magnitude $M_{0}$, and `$B$' is the
slope of the line in $\Delta$(colour)$/\Delta$(mag) as,
\[({\rm colour})=A+B({\rm mag}-M_0),\]
where mag and $M_0$ are absolute magnitudes.
The regression lines for the three clusters from E97 
(Cl~0016$+$16, Cl~0054$-$27 and Cl~0412$-$65)
on the $V_{555}-I_{814}$ vs. $I_{814}^T$ plane,
defined for spheroidal galaxies with apparent $I_{814}^T$ magnitudes brighter 
than 23\,mag, are taken from their paper with the zero-points colours
transformed to those at the absolute magnitude $I_{814}^T=-23$.
The slopes and zero-points of the {\it C-M\/} relations for the SED98 sample
have been taken from their paper as well.
We adopted their $V-K$ colours for 8 clusters between $z=0.461$ and $0.689$,
and $R-K$ colours for 3 clusters beyond that redshift, all of which
are referred as $blue-K$ colours in their paper.
Since they do not give the absolute colours of the {\it C-M\/} relation, 
we estimate the zero-points at $M_0$ from the absolute colours of
their no-evolution {\it C-M} relation at $M_0$ and the average colour
difference from it, both of which are given in the figures in SED98.
We note that the errors in zero-point `$A$' in Table~1 include only
the random errors of the data; the systematic errors of the
photometry are not included, except for the SED98 sample.
We discuss the systematic errors further in \S4.2.

For the most distant cluster CIG~J$0848+4453$ at $z=$1.273, we adopt
$R-K$ colours and $K$ magnitudes of the six spectroscopically-confirmed
cluster members for supplemental use, although morphology is unknown
(Stanford et al.\ 1997).  The {\it C-M\/} relation, however, cannot be
defined due to the paucity of the number of galaxies.  Thus, we measure
the average colour and the standard deviation of the colours using the 
error of each colour as a weight, and regard these results as the
zero-point colour and its error.  These numbers are also listed in
Table~1.

\section{Model}

\subsection{Evolution of the colour-magnitude relation}

The elliptical galaxy models we have used are essentially the same as
those built by KA97 (see also Kodama 1997 for details).  Following
Larson (1974) and Arimoto \& Yoshii (1987), we assume elliptical galaxy
formation occurs in a monolithic collapse accompanied by a galactic
wind. Star formation is burst-like with very short star-formation and
gas-infall time-scales (both are set to be 0.1~Gyr) followed by a
galactic wind which occurs less than 0.5~Gyr from the start of galaxy
formation.  Chemical evolution is taken into account consistently under
a well-mixed approximation.  The model is calibrated to the {\it C-M\/}
relation of Coma in the $V-K$ \vs $M_{V}$ plane (BLE92) at $z=0$,
either by changing the mean stellar
metallicity ({\it metallicity sequence}) or the age ({\it age sequence)}
of the galaxies.

We summarize now the small changes made on the KA97 models.  The
cosmological parameter $q_0$ was changed from the value $0.1$ used in
KA97 to $0.5$.  Following this change, we reconstructed the elliptical
galaxy models to have an age 12~Gyr instead of 15~Gyr to be consistent
with the shortening of the age of the Universe.  In this cosmology
($q_0=0.5$), the lookback time of 12~Gyr corresponds to redshift $z\sim
4.5$ (a lookback time of 15~Gyr corresponds to $z\sim 5.4$ with
$q_0=0.1$).  In this paper, we also consider the metallicity sequence
models for ages down to 9~Gyr ($z_f=1.2$).  Another small difference is
a change in the initial mass function slope.  Hereafter, we use
$x=1.10$ instead of the value $x=1.20$ adopted in KA97.  The reason is
the following. To construct the metallicity sequence model for
relatively younger ages, such as $9-10$~Gyr, the chemical yield is
insufficient for $x=1.20$, since the younger age population needs to be
made up of higher metallicity material to reproduce the same colours.
However, the analysis is almost totally independent of these small
changes (KA97). The properties of the metallicity sequence of
elliptical galaxies are summarized in Table~2,
where $M_{V}$, $M_{G}$, $T_{G}$, $t_{gw}$, $Z_g(t_{gw})$, and $M/L_B$
are total absolute magnitude in $V$-band, total stellar mass, galactic age,
time of the galactic wind, metallicity of the galactic gas at $t_{gw}$,
and mass-to-light ratio in $B$-band of the model galaxies at $z=0$,
respectively.
The luminosity-weighted mean stellar metallicities are given by two
definitions; $\langle\log Z/Z_{\odot}\rangle$ and
$\log \langle Z/Z_{\odot}\rangle$ (see KA97).
The comparison with
Coma ellipticals (BLE92) are presented in Fig.~2.  Finally, we apply an
aperture correction to the model when necessary, as described in
\S3.2.

\begin{table*}
\caption{Model sequence of elliptical galaxies at $z = 0$ ($T_G=$12 Gyr).}
\label{table-2}
\begin{center}
  \begin{tabular}{c|l|rrrrrrr}\hline\hline
&  $M_{V}({\rm mag})$       & $-$23.00  & $-$22.00  & $-$21.00  & $-$20.00  & $-$19.00  & $-$18.00  & $-$17.00 \\
&  $M_{G}(10^{9}M_{\odot})$ &   848  &   311  &   115  &   42.4  &   15.7  &   5.85  &   2.18 \\
&  $T_{G}$(Gyr) &  12.00  &  12.00  &  12.00  &  12.00  &  12.00  &  12.00  &  12.00 \\
&  $t_{gw}$(Gyr) &  0.353  &  0.256  &  0.199  &  0.158  &  0.128  &  0.106  &  0.089 \\
{\it metallicity} & $\langle\log Z/Z_{\odot}\rangle$ &  0.061  & $-$0.038  & $-$0.132  & $-$0.229 & $-$0.328  & $-$0.425  & $-$0.523 \\
{\it sequence} &  $\log \langle Z/Z_{\odot}\rangle$ & 0.202 & 0.094 & $-$0.005 & $-$0.101 & $-$0.196 & $-$0.290 & $-$0.382 \\
&  $Z_{g}(t_{gw})$ & 0.566 & 0.417 & 0.298 & 0.188 & 0.084 & $-$0.010 & $-$0.099 \\
&  $M / L_{B}$  &  8.830  &  8.002  &  7.113  &  6.459  &  5.978  &  5.451  &  5.049 \\
&  $U - V$      &  1.668  &  1.591  &  1.516  &  1.440  &  1.365  &  1.295  &  1.228 \\
&  $V - K$      &  3.355  &  3.274  &  3.194  &  3.113  &  3.033  &  2.952  &  2.871 \\ \hline\hline
  \end{tabular}\\
  \end{center}
\end{table*}

\begin{figure}
\begin{center}
  \leavevmode
  \epsfxsize 1.0\hsize
  \epsffile{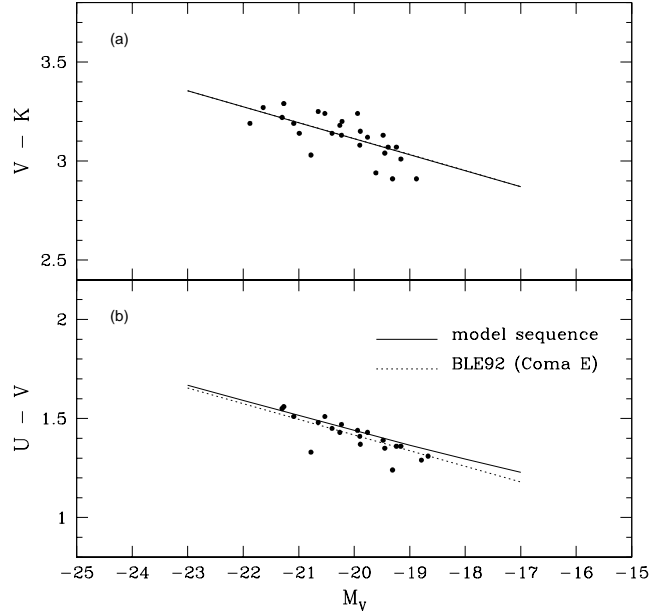}
\end{center}
\caption
{
The {\it C-M\/} relation for Coma ellipticals. The filled circles
represent the Coma ellipticals from Bower et al.\ (1992).  The solid
lines represent the loci of the metallicity sequence model with age
12~Gyr.
}
\label{fig:coma}
\end{figure}

We have simulated the evolution of the {\it C-M\/} relation as a
function of redshift both in the observer's frame and in the rest frame
for various photometric systems.  The response curves are taken from
Bessell (1990) for the Johnson $V$ and the Cousins $R$ and $I$ bands,
and from Bessell \& Brett (1988) for the $K$ band.  The {\it HST}
filter response functions for $V_{555}$, $R_{702}$, and $I_{814}$ are
taken from the Space Telescope Science Institute web page.
The photometric zero-points of the HST bands are taken from Holtzman \etal
(1995).

To make a precise comparison with the data, the slope of the {\it
C-M\/} relation at a given redshift is defined in the brightest 3\,mag
range of the model {\it C-M\/} relation by drawing a straight line
between the two model galaxies which have $M_{V}^{T}=-23$ and
$M_{V}^{T}=-20$ at $z=0$, respectively, which is comparable to the
definition of the slope for the observational data.  Note that the {\it
C-M\/} relation for the metallicity sequence is well represented by a
straight line within the redshift range under consideration, and the
above definition is quite reasonable.  The zero-point of the model {\it
C-M\/} relation is also defined at the absolute magnitude $M_{0}$, just
as was done for the observational data.

The full tables of the evolution of the colours of elliptical galaxies
in various combinations of age and metallicity and in various
photometric systems will be presented in a separate paper in the
supplement series (Kodama \& Arimoto 1997, in preparation).  The
machine readable version of the tables will also be provided upon
request.

\subsection{Aperture correction}

Since early-type galaxies usually have radial colour gradients (e.g.
Vader et al.\ 1988; Franx \& Illingworth 1990; Peletier et al.\ 1990a;
Peletier, Valentijn, \& Jameson 1990b; Balcells \& Peletier 1994),
their colour indices will depend on the adopted aperture within which
the galaxy light is integrated.  Barger et al.\ (1996, 1997) used
$5\,$arcsec aperture diameters to define their colour indices, which
correspond to $\sim 14h^{-1}$ kpc for $z=0.31$ and $\sim 18h^{-1}$ kpc
for $z=0.5$.  On the other hand, BLE92 used $11\,$arcsec apertures for
their Coma cluster galaxies, which corresponds to $5h^{-1}\,$kpc. Since
our model for elliptical galaxies is calibrated to match the {\it
C-M\/} relation of the Coma ellipticals in BLE92, we need to take into
account the aperture differences in order to accurately compare the
models with the observational data.

We apply the aperture correction to the models instead of to the
observational data of each cluster by constructing an aperture
corrected {\it C-M\/} relation of Coma and recalibrating the model to
match it (see KA97 for details).  We adopt the value $\Delta(V-K)/(\log
r/r_{e})=-0.16$ given by Peletier et al.\ (1990b) as a typical bright
elliptical galaxy colour gradient, where $r_e$ is the effective radius
of the galaxy. Gonz\'alez \& Gorgas (1996) recently suggested that
elliptical galaxies with stronger central Mg$_{2}$ indices show steeper
line-strength gradients. If we consider the fact that the central
Mg$_{2}$ strength strongly correlates with the velocity dispersion of
the galaxy (Bender, Burstein, \& Faber 1993; J$\o$rgensen, Franx, \&
Kj$\ae$rgaard 1996), then smaller ellipticals could have shallower
colour gradients. However, Gonz\'alez \& Gorgas (1996) failed to detect
significant correlations of the gradients with either absolute magnitude
or galaxy mass, thus we have adopted a constant colour gradient for 
all galaxy models. We use the $M_{K}$ \vs $\log r_{e}$ relation derived from
the Kormendy relation of Pahre, Djorgovski, \& Carvalho (1995) to apply
aperture corrections to the various galaxy sizes.  Using the colour
gradient and the above relation, as well as the de Vaucouleurs (1948)
$r^{1/4}$-law for the radial profile of ellipticals, we can reconstruct
the standard {\it C-M\/} relation at $z=0$ in the $V-K$ \vs $M_{V}$
diagram for any given aperture using the BLE92 {\it C-M\/} relation for
Coma.

\begin{figure}
\begin{center}
  \leavevmode
  \epsfxsize 1.0\hsize
  \epsffile{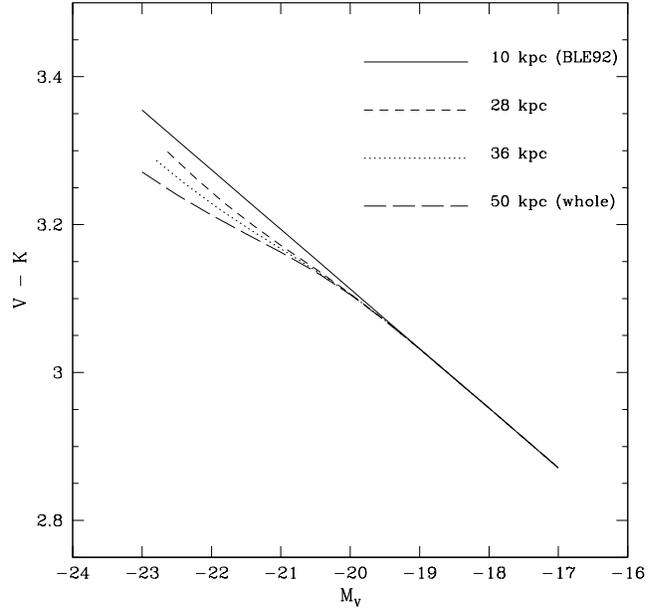}
\end{center}
\caption
{
 Aperture corrected {\it C-M\/} relations for Coma ellipticals.
 Three {\it C-M\/} relations for different apertures are constructed from the
 Bower et al. (1992) relation for Coma ellipticals,
 taking into account the aperture correction ($h=0.5$).
 The Bower et al. (1992) standard {\it C-M\/} relation is also shown.
}
\label{fig:aperture}
\end{figure}

The corrected {\it C-M\/} relations for Coma ellipticals are shown in
Fig.~\ref{fig:aperture} for three different physical aperture
diameters, 28, 36, and 50\,kpc.  The first two correspond to 5\,arcsec
at $z=0.31$ and at $z=0.5$, respectively ($h=0.5$), while the third
represents the relation for the whole galaxy. The {\it C-M\/} relation
of BLE92 is indicated by the solid line.  A bluing in the corrected
{\it C-M\/} relation compared to that of BLE92 is observed for bright
galaxies ($M_{V} \lsim -20\,$mag). This is because a larger aperture
covers more of the physical size of the galaxy, which contains more
blue light of presumablly metal-poor stars.
On the contrary, fainter galaxies have smaller physical
sizes, and hence a 10\,kpc aperture already covers most of the light of
the galaxy and needs little correction.  In spite of the bluing of the
{\it C-M\/} relations, the changes in the slope for 28 and 36\,kpc
apertures are tempered.  This is simply because the brighter galaxies
lose more light outside the restricted aperture. The amount of dimming
compared to the total magnitudes $M_{V}$ in BLE92 is larger for
brighter galaxies.  Thus, the aperture effect on the slopes of the {\it
C-M\/} relation is kept rather small. For any redshift under
consideration, the slope typically changes by only 0.01 and 0.02 per
mag for the 28\,kpc and 36\,kpc apertures compared to the 10\,kpc
aperture, respectively.  The effect on the zero-points of the {\it
C-M\/} relation is negligibly small in any case (less than 0.05\,mag).

When we compare the model to the photometric data for a given physical
aperture, we regard the reconstructed {\it C-M\/} relation for Coma as
the new standard relation, and the galaxy models are recalibrated so as
to reproduce it at $z=0$.  The adopted aperture for the correction of
the models are 28 kpc for AC~103, AC~114, and AC~118, and 36 kpc for Cl~
0024+16, and Cl~0016+16.  The $V_{555}-I_{814}$ colours for the three
clusters from E97 have the same 10\,kpc aperture as that of BLE92, and
hence no correction is applied.  For the rest of the clusters,
including those from SED98, 3C~324, and CIG~J0848+4453, the 28\,kpc
aperture model is applied, even though the SED98 sample and CIG~
J0848+4453 actually have $20-27\,$kpc and $17\,$kpc diameter apertures,
respectively.
Here the apertures for SED98 clusters are provided from S.A.~Stanford
(private communication).
This small aperture mismatch is negligible both in the
slopes and in the zero-points.  Note that the SED98 $K$ magnitudes are
total magnitudes and not magnitudes inside 28\,kpc, hence we apply a
slight correction to the 28\,kpc aperture model.
Since we lack aperture information on the 3C~324 data,
we adopt the 28\,kpc aperture model arbitrarily.
The corrections in the slope and zero-point are
likely to be small when compared to the observational uncertainties.

\section{Comparison}

The model predictions for the evolution of the {\it C-M\/} relation are
indicated by solid lines with changing redshift in
Fig.~\ref{fig:diagram} for the 6 clusters for which new {\it C-M\/}
regression lines were determined in this work.  The model is a
metallicity sequence with an age of 12~Gyr ($z_f=4.5$).  As is evident
from the figures, the observed {\it C-M\/} relations (dashed lines) are
well defined in clusters up to $z \simeq 1.2$, and the slopes evolve
almost in parallel, in good agreement with the model.  The zero-points
of the {\it C-M\/} relation for some of the clusters in Barger et
al.\ (1997) deviate from the model prediction by as much as 0.2\,mag.
This probably arises from the zero-point uncertainties of the data. We
discuss the zero-points in \S4.2.   The slopes of the
model {\it C-M\/} relations seem to globally match the observed data
very well.  This strongly suggests that there is no differential
evolution as a function of galaxy mass and that spheroidal galaxies in
clusters form universally at high redshift.  A detailed comparison with
the models for these and the other clusters are given below.

\subsection{Slopes}

\begin{figure*}
\begin{center}
  \leavevmode
  \epsfxsize 1.0\hsize
  \epsffile{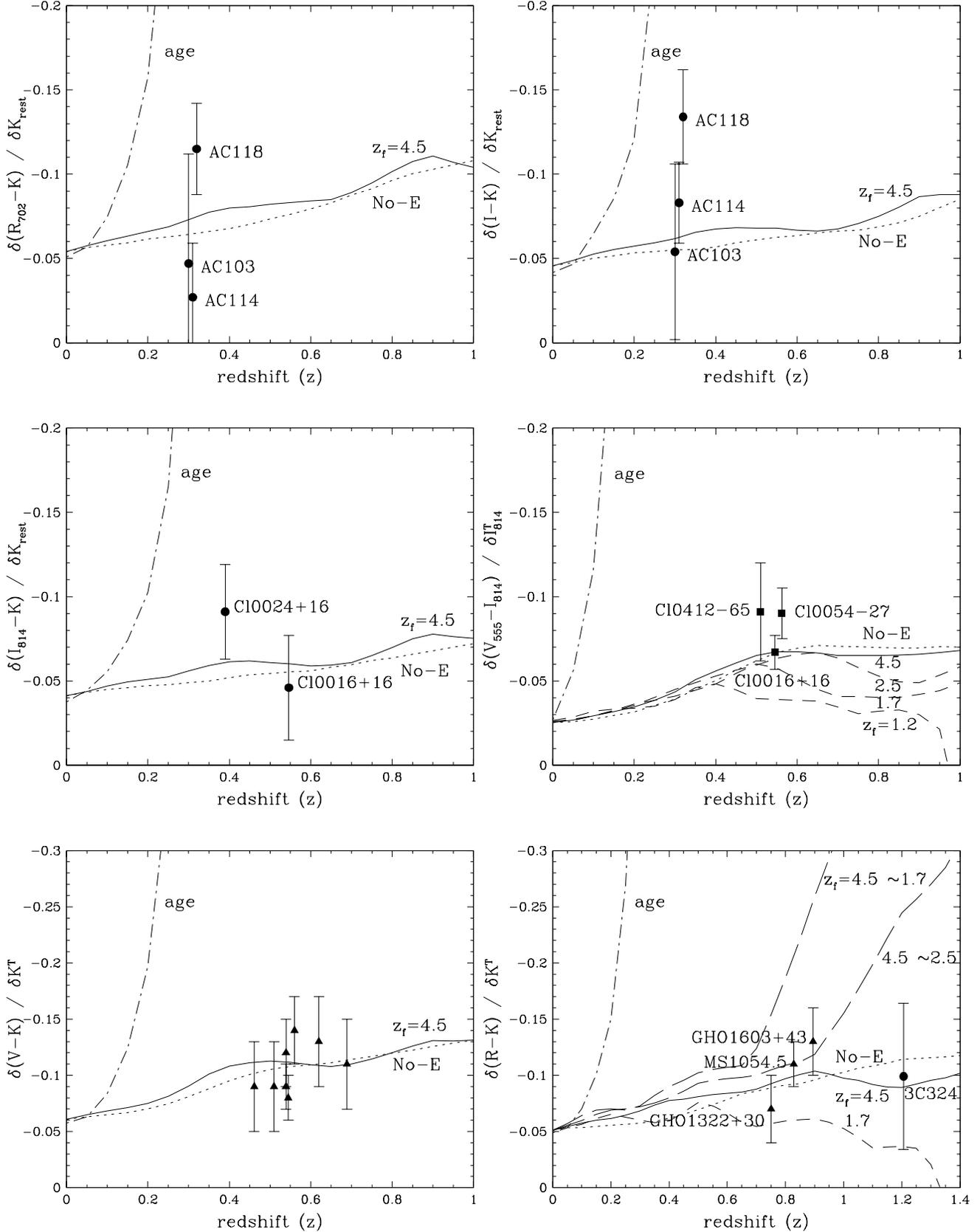}
\end{center}
\vspace{-0.7cm}
\caption
{
 Evolution of the slope of the {\it C-M\/} relation.
 The solid lines represent the metallicity sequence model with
 $z_f\simeq4.5$ ($T_{G}=12$ Gyr).
 The dashed lines correspond to models with different $z_f$s, as indicated
 in the figure.
 The long-dashed lines show the models that change $z_f$ as a function
 of galaxy luminosity for the brightest 3\,mag range
 (brighter galaxies have larger values of $z_f$).
 The dash-dot lines show the extreme age sequence models from KA97.
 The dotted lines represent the no-evolution prediction (No-E).
 Our original results from the Barger et al.\ (1997) sample are shown by
 filled circles. The E97 and SED98 samples are indicated by filled squares
 and filled triangles, respectively.
 In the bottom-left panel, each observational point corresponds
 respectively to following cluster in ascending redshift order:
 3C~295, Cl~$0412-65$, GHO~$1601+4253$ (down), MS~$0451.6-0306$ (up),
 Cl~$0016+16$, Cl~$0054-27$, 3C~220.1, and 3C~34.
}
\label{fig:slope}
\end{figure*}

The evolution of the slope of the {\it C-M\/} relation contains
critical information on the origin of the {\it C-M\/} relation itself,
i.e., which is the dominant factor controlling the systematic
difference in the photometric properties as a function of galaxy
luminosity. The evolution of the slope of the theoretical {\it C-M\/}
relation is compared with the observed data in Fig.~\ref{fig:slope}.
The solid lines show the standard metallicity sequence model with
$T_G=12$ Gyr ($z_f\simeq4.5$), the same as in Fig.~\ref{fig:diagram}.
Almost all of the clusters are consistent with the model prediction of
this single metallicity sequence with old age within 1.5$\sigma$
errors.  As is already mentioned in KA97, the metallicity sequence with an
old age keeps the slope of the {\it C-M\/} relation essentially unchanged,
although a slight steepening of the slope can be seen.  The dotted
lines indicate the no-evolution models estimated by simply redshifting
the $z=0$ models.  The no-evolution models also show a slight
steepening of the slope and are very close to the prediction of the
$z_f\simeq4.5$ metallicity sequence.  This indicates that the
steepening is simply due to the shift of corresponding wavelength
shortwards with redshift.  On the contrary, the dot-dashed lines
correspond to the age sequence model; i.e., the ages become younger
toward fainter galaxies along the {\it C-M\/} relation.  As is evident,
the pure age sequence is again absolutely rejected by all the clusters,
which strengthens the conclusion of KA97.

Note that only AC~118 could have a somewhat steeper slope than the
metallicity sequence model both in $R_{702}-K$ and $I-K$ (at the
1.5$\sigma$ and 2$\sigma$ levels, respectively), although still far
away from the age sequence. The significance of the difference is too
small to deserve further speculation.  In fact, SED98 also measured the
{\it C-M\/} slope for this cluster in a shorter wavelength colour
($G-K$, where $G$ denotes Gunn's $g$ filter) and found that it does not
have a significantly steeper slope than the no-evolution Coma model.

Since one of our goals is to constrain the formation epoch from the
{\it C-M\/} slope alone, we consider the younger metallicity sequence
models as well.  In the centre right panel of Fig.~\ref{fig:slope}, we
show three metallicity sequences with younger ages: $T_G=11\,$Gyr,
$T_G=10\,$Gyr and $T_G=9\,$Gyr at $z=0$, corresponding to $z_f\simeq
2.5$, 1.7, and 1.2, respectively.  In the bottom right panel we also
show the $z_f=1.7$ model.  The above models have been recalibrated to
reproduce the Coma {\it C-M\/} relation at the fixed age by adjusting
the mean stellar metallicity along the {\it C-M\/} relation.  However,
a strong constraint on the formation epoch cannot be made due to the
large uncertainties in the slope, which is defined for only 3\,mag from
the brightest end of the {\it C-M\/} relation.  Even the model with
$z_f=1.7$ cannot be rejected for many clusters.  In section 4.2 we show
that the analysis of the zero-points is much more effective in
providing direct information on the average formation epoch of the
galaxies than that of the slopes.

However, the evolution of the {\it C-M\/} slope provides information on
{\it relative\/} age variations of the early-type galaxies with
different luminosities. Thus we can constrain the maximum allowed age
difference along the {\it C-M\/} relation. To do so, we introduce now a
model in which age is differentially changed as a function of galaxy
luminosity.  The two long-dashed lines in the bottom right panel of
Fig.~\ref{fig:slope} represent models which allow an age difference of
1\,Gyr ($z_f=4.5$--$2.5$) or 2\,Gyr ($z_f=4.5$--$1.7$) in the 3\,mag
range of the {\it C-M\/} relation with the brightest galaxies having
the oldest age ($z_f=4.5$).  Note that the difference is much smaller
than for the age sequence, which requires a $\sim 7\,$Gyr difference.
Despite the large uncertainties in the measured slopes, the four most
distant clusters can clearly reject a 2\,Gyr age difference, since the
expected slopes at high redshift would be far too steep.  If our models
are correct, the age difference is unlikely to be more than 1\,Gyr, and
could be much smaller.

Our analysis of the evolution of the {\it C-M\/} slope indicates that
there is little differential evolution in early-type galaxies as a
function of galaxy luminosity, which strongly suggests that the bulk of
the stars in early-type galaxies in rich cluster environments are
coeval and form at a redshift well beyond unity.  Moreover, we find
little difference between the {\it C-M\/} slopes of different clusters,
most of which are consistent with a universal metallicity sequence with
old age.  This argues for a universal mechanism responsible for the
{\it C-M\/} relation of the early-type galaxies in these clusters.
Their photometric properties may be described primarily by only one
parameter: the mean stellar metallicity controlled by their mass.

\subsection{Zero-points}

\begin{figure*}
\begin{center}
  \leavevmode
  \epsfxsize 1.0\hsize
  \epsffile{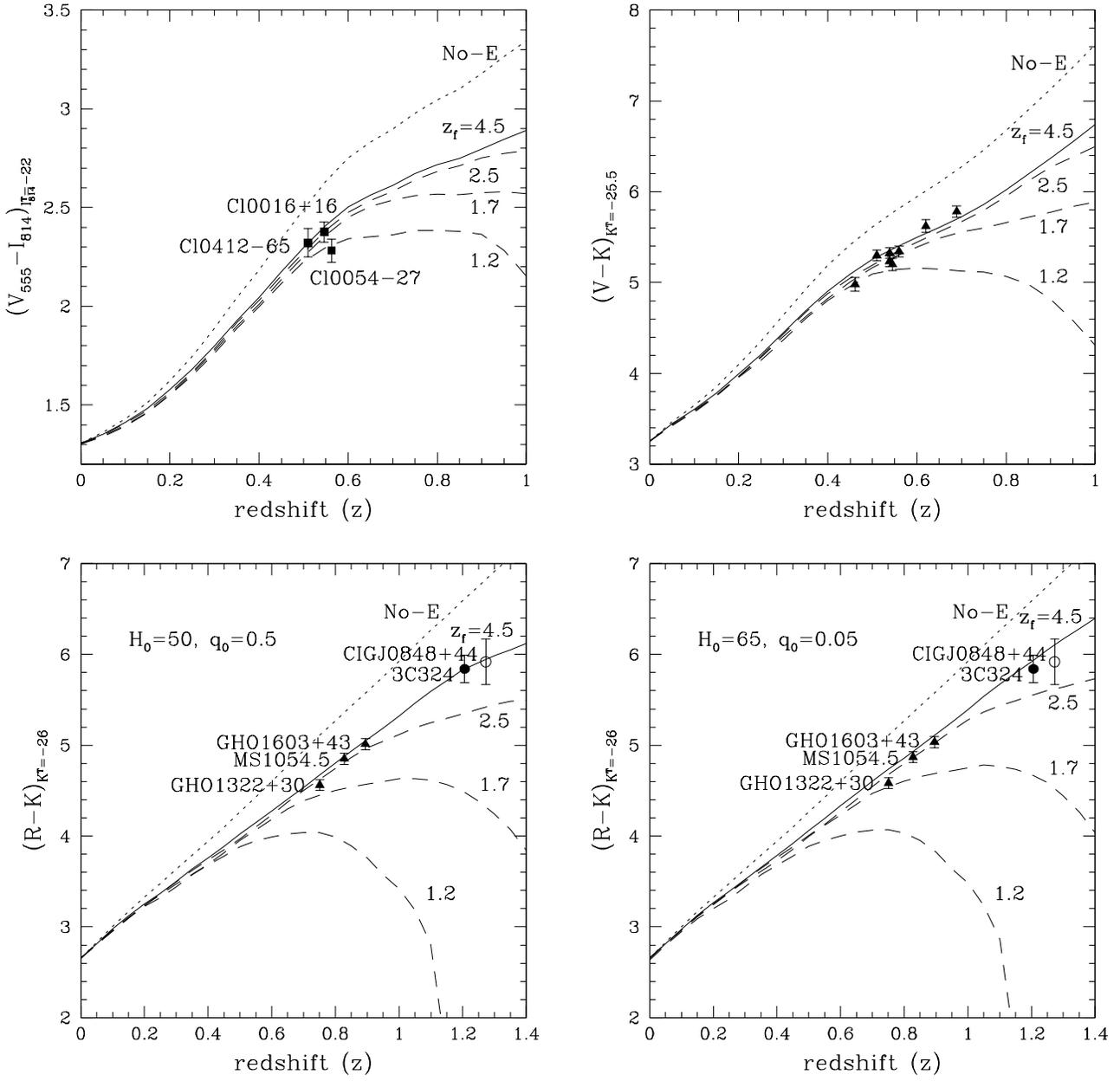}
\end{center}
\caption
{
 Evolution of the zero-point of the {\it C-M\/} relation.
 The dotted line indicates the no-evolution model (No-E).
 The solid lines represent the metallicity sequence model with
 $z_f\simeq4.5$ ($T_G=12$ Gyr).
 The dashed lines correspond to $z_f\simeq$ 2.5, 1.6, and 1.2,
 ($T_G=$11, 10, and 9 Gyr) from top to bottom, respectively.
 The observed data are shown by the symbols used in Figure~4.
 The average and the standard deviation of the galaxy colours 
 in CIG~J0848+44
 ($z=$1.273) are also shown by an open circle and an error bar.
 In the top-right panel, each observational point corresponds
 respectively to following cluster in ascending redshift order:
 3C~295, Cl~$0412-65$, GHO~$1601+4253$ (up), MS~$0451.6-0306$ (down),
 Cl~$0016+16$, Cl~$0054-27$, 3C~220.1, and 3C~34.
}
\label{fig:zerop}
\end{figure*}

The zero-points of the {\it C-M\/} relation contain direct information on
when the bulk of the stars formed in cluster early-type galaxies.
In Fig.~\ref{fig:zerop}, the zero-points `$A$' of the {\it C-M\/}
relation are compared with the models having various formation epochs.
The zero-points from Barger et al.\ (1997) are not shown because of
possible large uncertainties.  The dotted line indicates the
no-evolution model described above, and the four dashed lines
correspond to the metallicity sequence models with $T_{G} =$12, 11, 10,
and 9\,Gyr ($z_{f} =$ 4.5, 2.5, 1.6, and 1.2), from top to bottom,
respectively.  In the comparison, the systematic errors of the photometry
are also considered together with the random errors of the zero-points
shown in Table 1. In E97 $0.04\,$mag was given as the systematic
error.  For 3C~324, we conservatively choose 0.1 mag for the systematic
errors.  These numbers are added to the zero-point uncertainties of
`$A$' which only take into account random errors.  The systematic error
on the zero-points of the SED98 clusters are estimated by the authors
and are already included.  Note again that the zero-point error of CIG~
J$0848+4453$ is the standard deviation of the galaxy colours.

At redshifts below 0.6, the differences in the zero-points for the
model {\it C-M\/} sequences are too small to provide significant
constraints on the formation epoch.  Nevertheless, we can at least say
that the zero-points of these clusters are well below the no-evolution
prediction and are fully consistent with the metallicity sequence
models with old age.  Above that redshift, however, the zero-points get
progressively more sensitive to the differences in formation epoch, as
expected.  Between $z=$0.6 and 1.0, thanks to the small errors on the
zero-points, the formation epoch of early-type galaxies in SED98's 5
clusters can be constrained to $z_f>2$.  Beyond $z=1.0$, even a
$0.2-0.3\,$mag error is still small enough to discriminate between a
1\,Gyr difference in galaxy age.  In fact, as shown in the bottom left
panel of Fig.~\ref{fig:zerop}, the early-type galaxies in the two most
distant clusters 3C~324 and CIG~J$0848+4453$ should have a high
formation epoch, $z_f>3$.  

For $z>1$ the predicted colours depend slightly on the adopted
cosmology.  The bottom right panel in Fig.~\ref{fig:zerop} shows
alternative cosmology models with $H_{0}=65$ km s$^{-1}$ Mpc$^{-1}$ and
$q_{0}=0.05$.  Each observational zero-point has been adjusted
consistently with this cosmology due to a small shift of the position
on the {\it C-M\/} line that corresponds to $M_K=-26$ mag, where the
zero-point is defined, due to the change of distance modulus.  Note
that this correction is small, 0.02\,mag or less.  Although the latter
cosmology gives redder model colours and the constraint on the
formation epoch is slightly tempered, we can still say that the
formation epoch should be above redshift 2.  Even if the model
uncertainties were as large as 0.2 magnitudes, the result would not be
changed significantly.  The bulk of the stars in early-type galaxies in
cluster environments has to be formed at $z>2$ even for this
cosmology.

\section{Discussion and conclusions}

We have shown that the {\it C-M\/} relations of early type galaxies in
clusters with $0.31<z<1.27$ have slopes consistent with passively
evolving ellipticals of old age, with little differential evolution as
a function of galaxy luminosity.  The universality of the {\it C-M\/}
slope at high $z$ is in good agreement with the results found for
45 lower redshift clusters  ($0.02 < z < 0.18$) by  L\'opez-Cruz
(1997), suggesting that such universality is preserved over
substantial look-back-times.  Our model is fully consistent with the
observed slopes in the {\it C-M\/} diagram for almost all of the
clusters within one sigma errors (Kodama 1997).  The zero-points of the
{\it C-M\/} relation of high redshift clusters put strong constraints
on the formation redshift of the bulk of the stars in cluster
early-type galaxies: $z_{f} \geq 2-4$.

The following formation picture of cluster early-type galaxies,
especially the ellipticals, arises from our analysis:  Their stars are,
on the whole, very old, and the galaxies follow a universal function of
galaxy mass. The differences along the {\it C-M\/} sequence are
governed by the mean stellar metallicity.  Most of the early-type
galaxies in rich clusters form very quickly and effectively in the deep
potential well of the clusters at significantly high redshift, and
galactic wind feedback probably plays a key role in producing the {\it
C-M\/} relation.  Since elliptical galaxies are major constituents in
the core regions of galaxy clusters, the present study strongly
suggests that the central region of these clusters formed almost at the
same time in the early universe and by the same mechanism.

As shown in Fig.~\ref{fig:zerop}, the zero-point analysis of the {\it
C-M\/} relation provides very high sensitivity to the formation epoch
of the galaxies for clusters with $z>1$ even with moderately accurate
photometry (errors$\sim0.1\,$mag). Differences of $\sim1\,$Gyr are
easily measurable at such high redshifts.  Therefore, photometry of
early-type galaxies in clusters beyond $z\sim1$ can accurately
determine the formation epoch of these galaxies.
Furthermore, the slope of the {\it C-M\/} relation at redshifts beyond
$z\sim1$ is so sensitive to age difference
(as suggested by the long-dashed lines in the
bottom right panel of Fig.~\ref{fig:slope})
that it will allow us to investigate systematic difference of mean
stellar age as small as $\simeq$1~Gyr as a function of
galactic mass, if they are present.
We expect slope changes to be especially prominent when we approach the
star formation phase of the elliptical galaxies.

We have not discussed the colour scatter around the {\it C-M\/} relation
in this paper, but SED98 pointed out that the scatter dose not increase
with redshift out to $z\sim1$.
It would be extremely important to determine when the {\it C-M\/} relation
breaks down: i.e. when its scatter becomes very large.
This would indicate the very time when the ellipticals are forming.

Unfortunately, only a handfull of clusters have been found beyond
$z\sim1$.  Searching for such distant clusters and pushing the {\it
C-M\/} relation analysis towards higher redshifts must be a promising
strategy to determine the formation epoch of cluster ellipticals
globally and as a function of their mass.  The advent of large format
near-infrared arrays provides the means to carry out that search
efficiently.

We finish with a word of caution.  Our study is based on the monolithic
model for the formation of early-type galaxies.  The data we have
analyzed here is fully consistent with this model, but it might not be
the only solution.  In particular, an alternative scenario set in the
context of the hierarchical merging model of galaxy formation (e.g.
Kauffmann \& Charlot 1997) could give results that are also broadly
consistent with the available data, although it has not been fully
confirmed yet.  A detailed comparison of the observed properties of the
{\it C-M\/} relation in distant clusters with the predictions of
hierarchical models that properly take chemical evolution into account is
clearly needed.  It is especially important to test whether this type
of model can give the universal {\it C-M\/} relation for different
clusters locally and over a wide range of redshift,
since merging histories are likely to have varied from
cluster to cluster. Moreover it is not yet confirmed that larger ellipticals
form always from larger spirals already enriched in metal.
In any case, however, if we define the formation
epoch of the early-type galaxies as the time when the bulk of their
stars formed, the properties of the {\it C-M\/} relation place that
epoch well beyond $z=2$, independent of which formation picture is
correct. 

\begin{acknowledgements}

 We thank R.S.~Ellis, R.G.~Bower and S.A. Stanford for discussions.
 We are grateful to S.A.~Stanford for giving us continuous information
 of SED98 paper before publication.
 TK thanks the Japan Society for Promotion of Science (JSPS)
 for financially supporting his stay at the Institue of Astronomy, Cambridge,
 UK.
 NA thanks the JSPS for supporting his stay in Observatoire de Paris.
 This work was financially supported in part by a Grant-in-Aid for the
 Scientific Research (No.09640311) by the Japanese Ministry of Education,
 Culture, Sports and Science. AAS acknowledges generous financial support 
 from the Royal Society. 

\end{acknowledgements}


\begin{thebibliography}{}

 \bibitem{} Akritas, M. G., \& Bershady, M. A., 1996, ApJ, 470, 706
 \bibitem{} Arag\'on-Salamanca, A., Ellis, R. S., Couch, W. J.,
            \& Carter, D., 1993, MNRAS, 262, 764
 \bibitem{} Arimoto, N., \& Yoshii, Y., 1987, A\&A 173, 23
 \bibitem{} Balcells, M., \& Peletier, R. F., 1994, AJ, 107, 135
 \bibitem{} Barger, A. J., Arag\'on-Salamanca, A., Ellis, R. S.,
            Couch, W. J., Smail, I., \& Sharples, R. M., 1996, MNRAS, 279, 1
 \bibitem{barger97}
            Barger, A. J., Arag{\'o}n-Salamanca, A., Smail, I., Ellis,
            R. S., Couch, W. J., Dressler, A., Oemler, A., Poggianti,
            B. M., \& Sharples, R. M. 1997, ApJ, in press
 \bibitem{}  Bender, R., Burstein, D., \& Faber, S. M., 1993, ApJ, 411, 153
 \bibitem{} Bessell, M. S., \& Brett, J. M., 1988, PASP, 100, 1134
 \bibitem{} Bessell, M. S., 1990, PASP, 102, 1181
 \bibitem{} Bower R. G., Lucey J. R., \& Ellis R. S., 1992, MNRAS 254, 601
            (BLE92)
 \bibitem{} Burstein, D., \& Heiles, C., 1984, ApJS, 54, 33
 \bibitem{} Couch, W. J., Shanks, T., \& Pence, W. D., 1985, MNRAS, 213, 215
 \bibitem{couch97}
            Couch, W. J., Barger, A. J., Smail, I., Ellis, R. S., \&
            Sharples, R. M., 1997, ApJ, in press
 \bibitem{} de Vaucouleurs, G., 1948, Ann. d'Astrophys., 11, 247
 \bibitem{} Dickinson, M., 1996, in: Fresh Views of Elliptical Galaxies,
            eds. A. Buzzoni, A. Renzini, \& A. Serrano
            (ASP Conf. Ser. Vol. 86), p.~283
 \bibitem{} Dressler, A., Oemler, Jr., A., Couch, W. J., Smail, I.,
            Ellis, R. S., Barger, A., Butcher, H., Poggianti, B. M., \&
            Sharples, R. M., 1997, ApJ, in press
 \bibitem{} Ellis, R. S., Couch, W. J., MacLaren, I., \& Koo, D. C., 1985
            ApJ, MNRAS, 217, 239
 \bibitem{} Ellis, R. S., Smail, I., Dressler, A., Couch, W. C.,
            Oemler Jr, A., Butcher, H., \& Sharples, R. M., 1997,
            ApJ, 483, 582 (E97)
 \bibitem{} Franx, M., \& Illingworth, G. D., 1990, ApJ, 359, L41
 \bibitem{} Gonz\'alez, J. J., \& Gorgas, J. 1996, in: Fresh Views of
            Elliptical Galaxies, ed. A., Buzzoni, A. Renzini,
            \& A. Serrano (ASP Conf. Ser. Vol. 86), p.~225
 \bibitem{} Holtzman, J. A., Burrows, A. J., Casertano, S., Hester, J. J.,
            Trauger, J. T., Watson, A. M., \& Worthey, G., 1995, PASP, 107,
            1065
 \bibitem{} J$\o$rgensen, I., Franx, M., \& Kj$\ae$rgaard, P., 1996, MNRAS,
            280, 167
 \bibitem{} Kauffmann, G., \& Charlot, S., 1997, preprint, astro-ph/9704148
 \bibitem{} Kodama, T., 1997, Ph.D. Thesis, University of Tokyo
 \bibitem{} Kodama, T., Arimoto, N., 1997, A\&A, 320, 41 (KA97)
 \bibitem{} Larson, R. B., 1974, MNRAS, 166, 585
 \bibitem{} L\'opez-Cruz, O., 1997, Ph.D. Thesis, University of Toronto
 \bibitem{} Pahre, M. A., Djorgovski, S. G., \& de Carvalho, R. R., 1996, ApJ,
            456, L79
 \bibitem{} Peletier, R. F., Davies, R. L., Illingworth, G. D., Davis, L. E.,
            \& Cawson, M., 1990a, AJ, 100, 1091
 \bibitem{} Peletier, R. F., Valentijn, E. A., \& Jameson, R. F., 1990b,
            A\&A, 233, 62
 \bibitem{} Smail, I., Dressler, A., Couch, W. C., Ellis, R. S., 
            Oemler Jr, A., Butcher, H., \& Sharples, R. M., 1997,
            ApJS, 110, 481
 \bibitem{} Stanford, S. A., Eisenhardt, P. R. M., \& Dickinson, M., 1995,
            ApJ, 450, 512
 \bibitem{} Stanford, S. A., Eisenhardt, P. R. M., \& Dickinson, M., 1998,
            ApJ, 492, 461 (SED98)
 \bibitem{} Stanford, S. A., Elston, R., Eisenhardt, P. R. M., Spinrad, H.,
            Stern, D., \& Dey, A., 1997, AJ, 114, 2232
 \bibitem{} Vader, J. P., Vigroux, L., Lachi\`eze-Rey, M., \& Souviron, J.,
            1988, A\&A, 203, 217
 \bibitem{} Visvanathan, N., \& Sandage, A., 1977, ApJ, 216, 214

\end{thebibliography}
\end{document}